\documentclass[aps,prb,twocolumn,superscriptaddress]{revtex4}
\usepackage{graphicx}
\usepackage{amssymb}
\usepackage{amsmath}
\usepackage{graphicx}
\usepackage{epsfig}
\usepackage{color}
\usepackage{bm}

\usepackage{ulem}

\begin{document}

\title{Slave fermion interpretation of the pseudogap in doped Mott insulators}
\author{Zhuoqing Long}
\affiliation{Beijing National Laboratory for Condensed Matter Physics, Institute of Physics,
Chinese Academy of Science, Beijing 100190, China}
\affiliation{School of Physical Sciences, University of Chinese Academy of Sciences, Beijing 100190, China}
\author{Jiangfan Wang}
\affiliation{Beijing National Laboratory for Condensed Matter Physics, Institute of Physics,
	Chinese Academy of Science, Beijing 100190, China}
\author{Yi-feng Yang}
\email[]{yifeng@iphy.ac.cn}
\affiliation{Beijing National Laboratory for Condensed Matter Physics,  Institute of Physics, 
Chinese Academy of Science, Beijing 100190, China}
\affiliation{School of Physical Sciences, University of Chinese Academy of Sciences, Beijing 100190, China}
\affiliation{Songshan Lake Materials Laboratory, Dongguan, Guangdong 523808, China}
\date{\today}

\begin{abstract}
We apply the recently developed slave fermion approach to study the doped Mott insulator in the one-band Hubbard and Hubbard-Heisenberg models. Our results produce several subtle features in the electron spectra and confirm the key role of antiferromagnetic (AFM) correlations in the appearance of the pseudogap. Upon hole doping, the electron spectra exhibit a single peak near the Fermi energy in the local approximation of the Hubbard model where AFM correlations are not included. When AFM correlations are included through an explicit mean-field Heisenberg interaction, a second peak emerges at slightly lower energy and pushes the other peak to higher energy, so that a pseudogap emerges between the two peaks at small doping. Both peaks grow rapidly with increasing doping and eventually merge together, where the pseudogap no longer exists. Detailed analyses of the spectral evolution with doping and the strength of the Heisenberg interaction confirm that the lower-energy peak comes from a polaronic mechanism due to the holon-spinon interaction in the AFM-correlated background and the higher-energy peak arises from the holon hybridization to form the electron quasiparticles. Thus, the pseudogap arises from the interplay of the polaronic and hybridization mechanisms. Our results are in good agreement with previous numerical calculations using the dynamical mean-field theory and its cluster extensions, but give a clearer picture of the underlying physics. Our work provides a promising perspective for clarifying the nature of doped Mott insulators and may serve as a starting point for more elaborate investigations in the future.
\end{abstract}

\maketitle
\section{Introduction}
Despite tremendous efforts, the nature of the pseudogap in underdoped cuprates remains unresolved \cite{CupratesReview2015}. This mysterious phenomenon was first discovered by nuclear magnetic resonance (NMR) \cite{PG-NMR-1,PG-NMR-2,PG-NMR-3,PG-NMR-4}  and later confirmed by various other probes \cite{PseudogapReview1999}, manifested as a suppression of the density of states (DOS) far above the superconducting transition. In theory, cuprate physics is often described by doped Mott insulators \cite{MottInsulator2006Review}, but it has been questioned whether or not the pseudogap may involve extra factors like superconducting phase fluctuations or other symmetry-breaking mechanisms \cite{Pseudogap2015Review}. Nevertheless, a strong correlation effect must play an essential role \cite{MottInsulator2006Review, MottReview2010} and the doped Mott insulators have attracted many studies using various numerical methods such as the cluster extensions of the dynamical mean-field theory (DMFT) \cite{DMFTReview1996,clusterDMFTReview2005,HubbardModelNumericalReview2022} including the  dynamical cluster approximation (DCA) \cite{DCA1998,DCA2001} and the cellular DMFT (CDMFT) \cite{CDMFT2001}. These methods find indeed a pseudogap in the quasiparticle spectra of the one-band Hubbard model  \cite{Huscroft2001DCApseudogap,Macridin2006DCApseudogap,Kyung2006CDMFTpseudogap,Stanescu2006CDMFTzeropole,Sakai2009CDMFTzeropole,Sakai2010CDMFTzeropole,Vidhyadhiraja2009DCA_PGQCP,Ferrero2009momentumselective,Werner2009momentumselective,Gull2009momentumselective,Gull2010momentumselective,Tong2009CDMFT,Sordi2010Widomline,Sordi2012WidomlinePRL,Sordi2012WidomlineSciRep,FluctuationDiagnostics2015DCA,Wu2018FSTopology,Wu2020VanHove} and ascribe it to short-range antiferromagnetic (AFM) correlations \cite{Huscroft2001DCApseudogap,Macridin2006DCApseudogap,Kyung2006CDMFTpseudogap}. Explanations have been proposed from various  different aspects such as the reconstructions of pole-zero structure of the Green's function \cite{Stanescu2006CDMFTzeropole,Sakai2009CDMFTzeropole,Sakai2010CDMFTzeropole}, the ``momentum-selective'' Mott transition \cite{Ferrero2009momentumselective,Werner2009momentumselective,Gull2009momentumselective,Gull2010momentumselective}, and the organizing principle of the Widom line \cite{Sordi2010Widomline,Sordi2012WidomlinePRL,Sordi2012WidomlineSciRep}. Recent numerical works have provided further evidence for the key role of AFM correlations in the pseudogap phenomenon \cite{FluctuationDiagnostics2015DCA,FluctuationDiagnostics2017diagmc,Pseudogap2022diagMC}, but a thorough theoretical understanding is not yet available.

In this work, we apply our recently developed slave fermion approach \cite{SlaveFermion2022HalfFilling} to study the doped Mott insulators in the one-band Hubbard and Hubbard-Heisenberg \cite{Hubbard-Heisenberg-1,Hubbard-Heisenberg-2} models and explore the long-standing pseudogap enigma from a different perspective. This approach splits electrons into auxiliary fermionic doublons and holons carrying the charge degree of freedom and bosonic spinons carrying the spin degree of freedom \cite{Yoshioka1989SlaveFermion,Han2016SlaveFermion,Han2019SlaveFermion,SlaveFermion2022HalfFilling}. A fermionic auxiliary field is then introduced to decouple the kinetic term, and the spectra are calculated under the self-consistent one-loop local approximation. We find the pseudogap phenomenon is indeed closely associated with AFM correlations. For the Hubbard model, in which AFM correlations are not included as in DMFT, the hole doping yields a single sharp quasiparticle peak near the Fermi energy due to the charge Kondo effect of holons to form the electron quasiparticles \cite{SlaveFermion2022HalfFilling} and the calculated resistivity resembles that from the single-impurity DMFT. Once AFM correlations are taken into account through an explicit mean-field Heisenberg term as in the Hubbard-Heisenberg model, a second peak emerges below the Fermi energy due to the spin-polaronic mechanism \cite{Han2016SlaveFermion,Han2019SlaveFermion,SpinPolaron1988,SpinPolaron1989,SpinPolaron1991,SpinPolaron1992}, giving rise to the pseudogap around the Fermi energy in the electron spectra for small doping consistent with CDMFT calculations of the Hubbard model, which contain the effect of short-range AFM correlations. Thus, the pseudogap arises from the competition of quasiparticle and polaron formations. This provides a clearer physical picture of the pseudogap, and may serve as a starting point for further investigations.

\section{Method}
We start with the following model Hamiltonian on the square lattice:
\begin{eqnarray}
H=&-&\sum_{ij\sigma}t_{ij}c_{i\sigma}^\dagger c_{j\sigma}-\mu\sum_{i\sigma}c_{i\sigma}^\dagger c_{i\sigma}+J_H\sum_{\langle ij\rangle}\bm S_i\cdot\bm S_j\notag\\
&+&U\sum_{i}\left( n_{i\uparrow}-\frac12\right) \left( n_{i\downarrow}-\frac12\right),
\label{Hubbard model}
\end{eqnarray}
where the spin interaction $J_H$ is explicitly included because it cannot be generated automatically in the local approximation to be used in this work. The model is therefore also called the Hubbard-Heisenberg model. We will only consider the nearest-neighbor hopping and set $t=1/4$ so that the half bandwidth $D=4t=1$. The Hubbard interaction is set to $U=3.0$ to get a Mott insulator at half filling. The chemical potential $\mu$ will be tuned to control the hole doping.

In the slave fermion method \cite{Yoshioka1989SlaveFermion,Han2016SlaveFermion,Han2019SlaveFermion,SlaveFermion2022HalfFilling}, the physical electron operator is written as $c_{i\sigma}=h^\dagger_is_{i\sigma}+\sigma s^\dagger_{i,-\sigma}d_i$, where $d_i$ and $h_i$ are fermionic  doublon and holon operators, respectively, and $s_{i\sigma}$ are bosonic spinons, with the local constraint $Q_i\equiv h^\dagger_ih_i+d^\dagger_id_i+\sum_\sigma s^\dagger_{i\sigma}s_{i\sigma}=1$. In this representation, the spin interaction may be approximated by the Schwinger boson mean-field Hamiltonian \cite{Sachdev1991Sp(N)}: $J_H\sum_{\langle ij\rangle}\bm S_i\cdot\bm S_j \rightarrow H_{\text{spin}}^{\text{MF}}=\Delta\sum_{\bm k}\eta_{\bm k}(s^\dagger_{\bm k\uparrow}s^\dagger_{-\bm k,\downarrow}+\text{H.c.})$, where $\Delta=J_H|A|/(2t)$ is the ratio between spinon and electron bare bandwidths, $A=\sum_\sigma\langle\sigma s_{j,\sigma}s_{i,-\sigma}\rangle$ reflects AFM correlations between nearest-neighbor spins, and $\eta_{\bm k}=2t[\sin(k_x)+\sin(k_y)]$. The Hubbard term now takes a quadratic form of doublons and holons, but the hopping becomes quartic and shall be decoupled via the Hubbard-Stratonovich transformation by introducing a fermionic auxiliary field $\chi_{i\sigma}$.

\begin{figure}
\includegraphics[width=9cm]{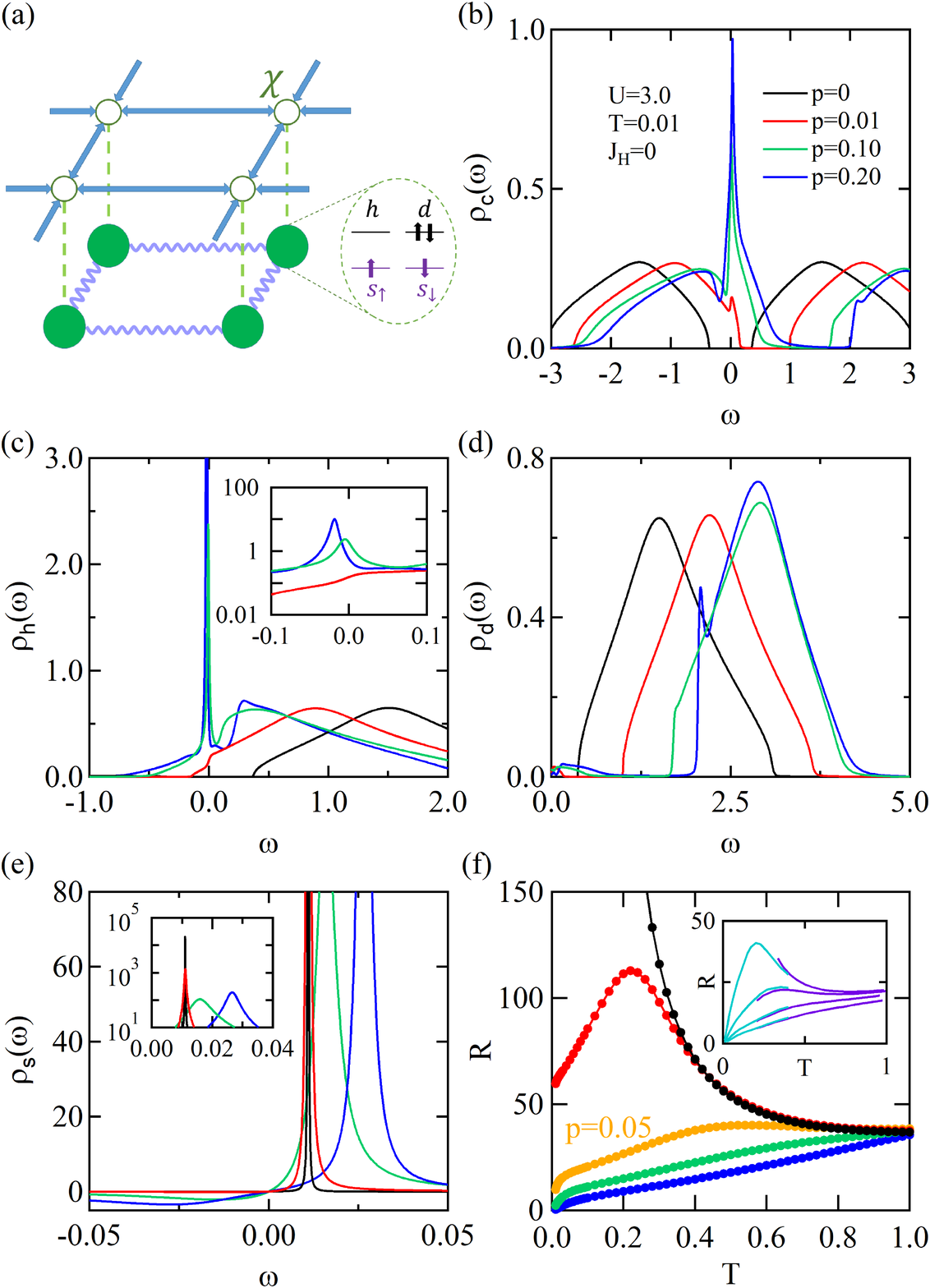}
\caption{(a) Illustration of the auxiliary fields showing the effective hopping of the fermionic $\chi$ field, the AFM correlations between neighboring spinons, and the local three-particle vertex as reflected in the effective Lagrangian. (b)-(e) Doping dependence of the densities of states of the electron ($\rho_c$), holon ($\rho_h$), doublon ($\rho_d$), and spinon ($\rho_s$) at $T=0.01$ and $U=3$. The insets of (c) and (e) show the holon and spinon spectra around zero energy. (f) Doping dependence of the calculated resistivity as a function of temperature. The result for $p=0.05$ is also shown for comparison. The inset reproduces the DMFT resistivity at $U=2.5$, obtained by using the numerical renormalization group (NRG) impurity solver for $p=0.02$, $0.05$, $0.10$, $0.15$ (top to bottom, cyan lines) \cite{Mazitov2022SquareLatticeResistivity}, and by using the continuous-time quantum Monte Carlo (CTQMC) and NRG impurity solvers for $p=0.00$, $0.05$, $0.10$,  $0.15$ (top to bottom, purple lines) \cite{Vucicevic2019SquareLatticeResistivity}. }
\label{fig1}
\end{figure}

Interestingly, we find a redundancy when dealing with the chemical potential term. Because of the constraint $Q_i=1$, the electron occupation operator obeys the equality: $n_i=\sum_{\sigma} c^\dagger_{i\sigma}c_{i\sigma}=2d_i^\dagger d_i+\sum_\sigma s^\dagger_{i\sigma}s_{i\sigma}=d_i^\dagger d_i-h^\dagger_ih_i+1$. Thus, the chemical potential can be treated either as the kinetic term of the physical electrons and then decoupled, or as a fictitious field splitting the doublon and holon levels. In principle, they should yield the same results if the model is solved exactly and the electron spectra are well reproduced by the slave particles. This is, however, spoiled because of the approximation. We find it very useful to take advantage of this redundancy and split the chemical potential into two terms, $-\mu\sum_{i\sigma}c_{i\sigma}^\dagger c_{i\sigma} \rightarrow -\sum_{ij\sigma}\mu_1\delta_{ij}c_{i\sigma}^\dagger c_{j\sigma}-\sum_i\mu_2(d_i^\dagger d_i-h^\dagger_ih_i+1)$, with $\mu=\mu_1+\mu_2$. This gives a free parameter to enforce the relationship $\langle\sum_{\sigma} c^\dagger_{i\sigma}c_{i\sigma}\rangle=\langle d_i^\dagger d_i-h^\dagger_ih_i+1\rangle$ under approximation. Similar redundancy also appears in the Kotliar-Ruckenstein slave boson method \cite{Kotliar1986SlaveBoson} and the $Z_2$ slave spin method \cite{de'Medici2005SlaveSpin,de'Medici2010SlaveSpin}. The final effective Lagrangian reads
\begin{eqnarray}
\mathcal{L}&=&\sum_i\left(\bar d_i \partial_\tau d_i+\bar h_i\partial_\tau h_i+\sum_\sigma\bar s_{i\sigma}\partial_\tau s_{i\sigma}\right)\notag\\
&-&\sum_{ij\sigma}\mathcal G_{ij}^{-1}\bar\chi_{i\sigma}\chi_{j\sigma}+\sum_{i\sigma}\left[\bar\chi_{i\sigma}(\bar h_{i}s_{i\sigma}+\sigma \bar s_{i,-\sigma}d_{i})+\text{H.c.}\right],\notag\\
&+&\left(\frac{U}{2}-\mu_2\right)\sum_i\bar d_id_i+\left(\frac{U}{2}+\mu_2\right)\sum_i\bar h_i h_i\notag\\
&+&\sum_i\lambda_i\left(Q_i-1\right)+H_{\text{spin}}^{\text{MF}},
\label{effective action}
\end{eqnarray} 
where $\lambda_i$ is the Lagrange multiplier for the local constraint and $\mathcal G_{ij}=-(t_{ij}+\mu_1\delta_{ij})$. Now, the slave particles $h_i$, $s_{i\sigma}$, and $d_i$ all sit on their own sites and are coupled with $\chi_{i\sigma}$, as illustrated in Fig. \ref{fig1}(a). The hopping (blue arrows) is entirely carried by the $\chi$-field, while the spin interaction offers another channel for spatial correlation (purple wavy line). We will see that the existence of both channels is essential for the pseudogap to emerge.

The model is then solved with the self-consistent one-loop approximation using the self-energy equations:
\begin{eqnarray}
&&\Sigma_{\chi}(\text{i}\omega_n)=\frac{1}{\beta}\sum_mG_s(\text{i}\nu_m)[G_h(\text{i}\omega_{m-n})-G_d(\text{i}\omega_{m+n})],\notag\\
&&\Sigma_s(\text{i}\nu_m)=\dfrac{1}{\beta }\sum_{n}G_\chi(\text{i}\omega_n)\left[G_d(\text{i}\omega_{n+m})-G_h(\text{i}\omega_{m-n})  \right],\notag\\
&&\Sigma_{h}(\text{i}\omega_n)=\frac{2}{\beta}\sum_mG_s(\text{i}\nu_m)G_\chi(\text{i}\omega_{m-n}),\notag\\
&&\Sigma_{d}(\text{i}\omega_n)=-\frac{2}{\beta}\sum_mG_s(\text{i}\nu_m)G_\chi(\text{i}\omega_{n-m}),
\label{eq:SelfE}
\end{eqnarray}
where $\text{i}\omega$ ($\text{i}\nu$) are the fermionic (bosonic) Matsubara frequencies, $\epsilon_{\bm k}$ is the bare dispersion of electrons, $\Sigma$ are the local self-energies, and $G$ are the full local Green's functions of the auxiliary fields. The electron Green's function is then given by 
\begin{equation}
G_c(\bm k,\text{i}\omega_n)=\dfrac{\Sigma_\chi(\text{i}\omega_n)}{1-(\epsilon_{\bm k}-\mu_1)\Sigma_\chi(\text{i}\omega_n)}.
\label{eq:Gc}
\end{equation}
To simplify the calculations, we have ignored the momentum dependency in the self-energies of auxiliary particles and replaced the Lagrange multipliers $\lambda_i$  with their mean-field value $\lambda$. The Green's functions and self-energies are then determined self-consistently. Note that all our calculations are performed in real frequency. For each step of iterations, we tune $\lambda$, $\mu_1$, and $\mu_2$ to enforce the three conditions $\sum_i\langle Q_i\rangle/N=1$, $\sum_{i\sigma}\langle c^\dagger_{i\sigma}c_{i\sigma}\rangle/N=1-p$, and $\sum_{i}\langle h^\dagger_{i}h_{i}-d^\dagger_{i}d_{i}\rangle/N=p$, where $N$ is the number of the lattice site and $p$ is the hole doping level. For $J_H\neq0$, the AFM mean-field parameter $\Delta$ is also determined self-consistently in each step. More details on the self-consistent equations and the numerical calculations are given in the Appendix.

\section{Results and discussion}
We first solve the self-consistent equations for $J_H=0$ containing no explicit AFM correlations. Figures \ref{fig1}(b)-(e) plot the obtained spectra of the electron and slave particles. At half filling ($p=0$, $\mu_1=\mu_2=0$), the electron spectra show two broad Hubbard bands and a finite Mott gap. Correspondingly, the holon and doublon spectra contain a single broad peak around the bare energy $\omega=U/2+\lambda$. A small hole doping such as $p=0.01$ shifts the lower Hubbard band of the electron spectra in Fig. \ref{fig1}(b) to zero energy and yields a quasiparticle peak signaling the doping-driven insulator-to-metal transition. The holons and doublons are no longer degenerate due to the chemical potential $\mu_2$, which pushes the holon spectra toward $\omega=0$ as shown in Fig. \ref{fig1}(c) and the doublon spectra to higher energies as shown in Fig. \ref{fig1}(d). As the holon band touches the Fermi energy, a sharp peak emerges around $\omega=0$ and has been attributed to a dynamic charge Kondo effect \cite{SlaveFermion2022HalfFilling} due to the three-particle vertex in Eq. (\ref{effective action}) that couples the holon, spinon, and $\chi$ field, with the spinon serving as the hybridization field. The sharp peak is  then nothing but a holon hybridization peak, which in turn leads to the quasiparticle peak in the electron spectra shown in Fig. \ref{fig1}(b). For large doping such as $p=0.2$, another sharp peak emerges at the lower edge of the doublon's broad band. Correspondingly, a small peak appears also in the high-energy part (upper Hubbard band) of the electron spectra in Fig. \ref{fig1}(b). This is probably associated with the sharp peak in the holon spectra and arises from quasiparticle formation, which necessarily involves the doublon. The spinon spectra are shown in Fig. \ref{fig1}(e) and exhibit a peak for all doping, which is, however, strongly damped by coupling to holons. The spinon number is mainly contributed by the broad background at finite $p$ while by the sharp peak at half filling. For comparison, we have calculated the resistivity in Fig. \ref{fig1}(f) using the same formula as in DMFT \cite{DMFTReview1996}, and the overall features are also very similar to those from DMFT \cite{DMFTReview1996,Mazitov2022SquareLatticeResistivity,Vucicevic2019SquareLatticeResistivity}. For small doping close to the Mott insulator such as $p=0.01$, the resistivity first grows rapidly as $T$ decreases and behaves like an insulator, but then falls and exhibits a broad peak at low temperatures. For larger doping, the resistivity turns metallic in the whole temperature window.

\begin{figure}
	\includegraphics[width=9cm]{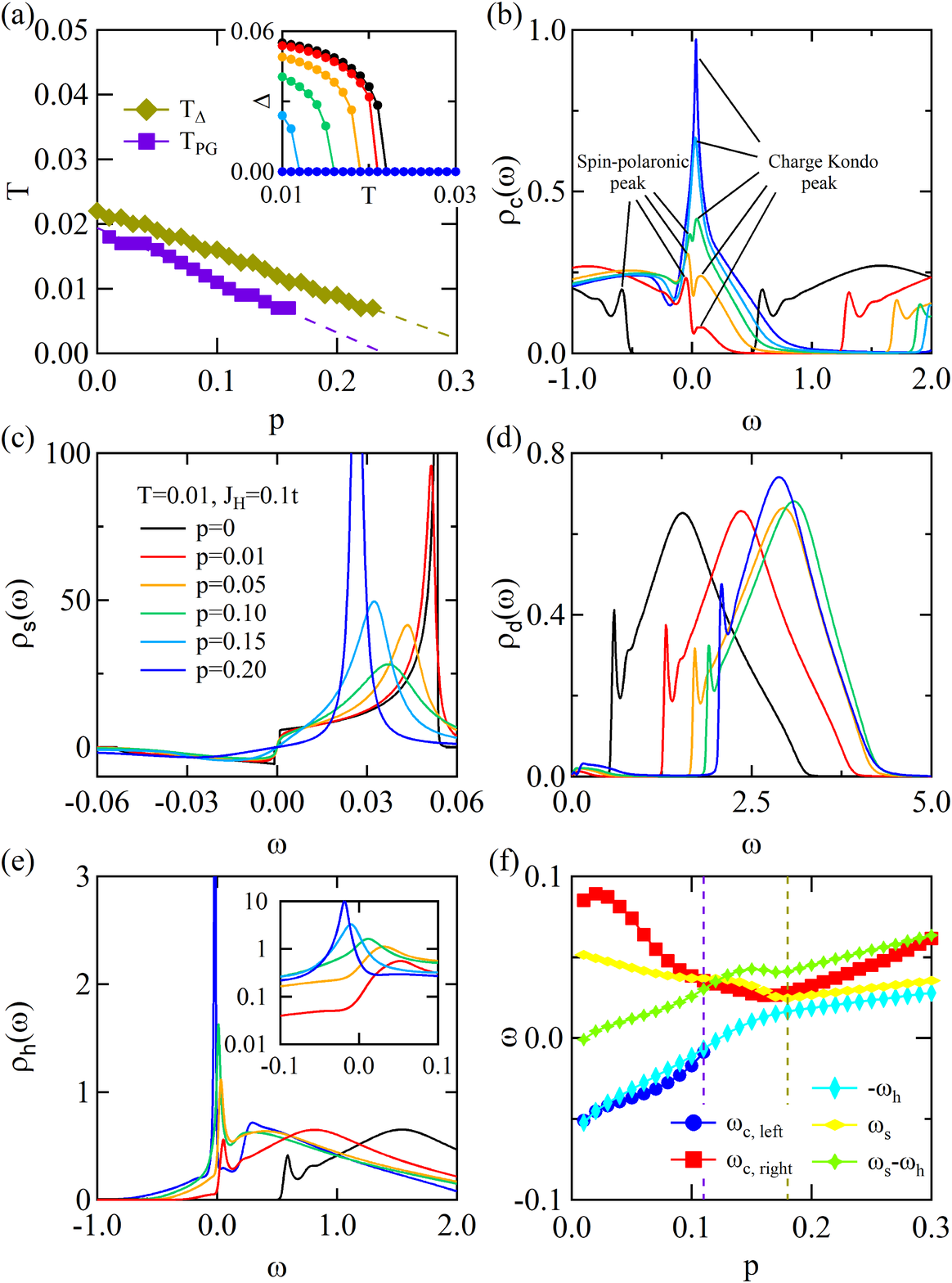}
	\caption{(a) Doping dependence of the onset temperature of the pseudogap $T_\text{PG}$ and AFM correlations $T_\Delta$ for $J_H=0.1t$ and $U=3$. The inset shows the temperature dependence of $\Delta$ at different dopings. (b)-(e) Doping dependence of the densities of states of the electron ($\rho_c$), spinon ($\rho_s$), doublon ($\rho_d$), and holon ($\rho_h$) at $T=0.01$. For clarity, the $p=0.15$ curve is not shown in (d) and the main panel of (e). The inset of (e) shows the holon spectra near zero energy. (f) Comparison of the various peak positions as a function of hole doping for the left and right peaks around the pseudogap on electron spectra ($\omega_{c,\text{left}}$ and $\omega_{c,\text{right}}$), the spinon peak $\omega_s$, the holon peak $\omega_h$, and their difference $\omega_s-\omega_h$. The two vertical lines denote the doping $p\approx0.11$ where the pseudogap vanishes and $p\approx0.18$ where the AFM correlations vanish at $T=0.01$.}
	\label{fig2}
\end{figure}

So far we have considered the local approximation of the Hubbard model and ignored AFM correlations. No pseudogap is seen in the electron spectra. Next, we restore the AFM correlations by including an extra Heisenberg mean-field term and perform the calculations for $J_H=0.1t$, a value specially chosen to get a phase diagram of similar critical $p$ as in cuprates. The inset of Fig. \ref{fig2}(a) shows the calculated AFM correlation parameter $\Delta$ at different dopings. As expected, $\Delta$ decreases gradually with increasing temperature and finally diminishes. Its onset temperature $T_\Delta$ is plotted in Fig. \ref{fig2}(a), which decreases with increasing doping and extrapolates to zero for $p\gtrsim0.3$. 

Figure \ref{fig2}(b) shows the resulting electron spectra for $T=0.01$, at which temperature AFM correlations only exist for $p<0.18$. The Mott gap at half filling is enlarged due to the doublon-holon binding \cite{Castellani1979DHBinding, Kaplan1982DHBinding, Capello2005DHBinding, Yokoyama2006DHBinding, MottReview2010, Zhou2014DHBinding, Sato2014DHBinding, 	Prelovsek2015DHBinding, Han2016SlaveFermion, Han2019SlaveFermion, Zhou2020DHBinding, Terashige2019DHbindingExperiment} caused by a $d_ih_j$-like interaction after integrating out the dispersive spinons \cite{SlaveFermion2022HalfFilling}. As shown in Figs. \ref{fig2}(d) and \ref{fig2}(e), sharp inner peaks emerge at the edges of the Mott gap in the electron spectra and the lower edge of the broad holon and doublon spectra. As discussed already in our previous work \cite{SlaveFermion2022HalfFilling}, they are recognized as spin-polaronic peaks due to the coupling of doublon or holon with the correlated AFM background. Similar peaks have also been obtained in the self-consistent Born approximation calculation of the Hubbard model at half filling \cite{Han2016SlaveFermion,Han2019SlaveFermion} and the $t$-$J$ model \cite{SpinPolaron1988,SpinPolaron1989,SpinPolaron1991,SpinPolaron1992}.

\begin{figure}
	\includegraphics[width=9cm]{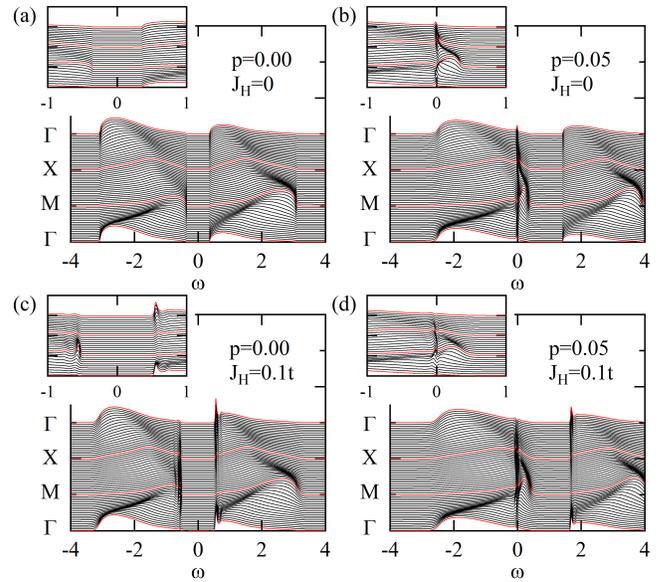}
	\caption{Comparison of the momentum-dependent electron spectra for (a) $p=0$ and $J_H=0$, (b) $p=0.05$ and $J_H=0$, (c) $p=0$ and $J_H=0.1t$, and (d) $p=0.05$ and $J_H=0.1t$. The red lines mark the spectra at $\Gamma\equiv(0,0)$, $X\equiv(\pi,0)$, and $M\equiv(\pi,\pi)$.}
	\label{fig3}
\end{figure}

Upon hole doping, the lower Hubbard band is expected to move toward zero energy. But quite amazingly, two peaks, instead of one,  emerge and give rise to a pseudogap or a dip around $\omega=0$. Including a negative next-nearest-neighbor hopping $t'$ seems to enhance slightly the pseudogap \cite{NNNhoping_ED, NNNhoping_CPT}. With increasing hole doping, both peaks are enhanced but the right one at higher energy grows more rapidly.  For large $p$ ($>0.11$), the lower-energy left one is absorbed and two peaks merge together. When $\Delta$ vanishes for $p>0.18$, the low-energy spectra behave as those for $J_H=0$ in Fig. \ref{fig1}(b). The overall doping dependence agrees well with that from CDMFT \cite{Kyung2006CDMFTpseudogap}. The pseudogap is gradually suppressed at high temperatures and its onset temperature $T_{\text{PG}}$ is also plotted in Fig. \ref{fig2}(a) for comparison. Interestingly, we see it is always smaller than the AFM correlation temperature $T_{\Delta}$, varies roughly linearly with doping, and extrapolates to zero for $p\gtrsim0.23$.

For completeness, the slave particle spectra are shown in Figs. \ref{fig2}(c)-(e). AFM correlations carried by the Heisenberg mean-field term change the spinon spectra, which exhibit a sharp peak roughly at $\omega_s\approx \Delta$ and a sharp spin gap near $\omega=0$ for $p=0$. As the doping increases, the peak moves to lower energy and gets broadened for small $p$, but grows again for larger $p$ as $\Delta$ goes to zero, recovering those for $J_H=0$ in Fig. \ref{fig1}(e). The doublon spectra shown in Fig. \ref{fig2}(d) contain a sharp inner peak compared to that for $J_H=0$. Since this peak is present all along, it should be attributed to the polaronic mechanism at small doping, and to the quasiparticle formation at large doping. The holon spectra are shown in Fig. \ref{fig2}(e), and quite unexpectedly, contain only one peak near $\omega=0$, unlike the electron spectra. This peak exists already at zero doping, moves gradually to the Fermi energy with increasing hole doping, and eventually behaves as those at $J_H=0$, indicating again the evolution from polaronic to hybridization origin. 

It may be informative to compare the doping dependence of all these peak positions. As shown in Fig. \ref{fig2}(f), the left peak on the electron spectra only exists for small doping and follows closely the holon peak. It may thus be identified to arise from the polaronic mechanism. The right peak behaves more complicatedly. For large doping where $\Delta=0$, it follows roughly $\omega_s-\omega_h$, where $\omega_s$ ($\omega_h$) is the position of the spinon (holon) peak. We attribute it to the hybridization peak arising from the convolution of the holon and spinon spectra in calculating the electron spectra. But at small doping, where $\Delta\neq0$ and the spinons have a finite bandwidth, a simple relation no longer exists and the right peak is pushed to higher energy by the increasing AFM correlations ($\Delta$) with decreasing doping.

To further clarify the origin and the relation of these subtle features at small doping, we plot the momentum-dependent spectral functions in Fig. \ref{fig3}. Due to the local approximation of the self-energies, we cannot study the nodal-antinodal dichotomy in the spectra \cite{PG-ARPES-1,PG-ARPES-2,PG-ARPES-3,PG-ARPES-4,PseudogapARPESReview2014}. Figure \ref{fig3}(a) shows the electron spectra at half filling with $J_H=0$. A clear Mott gap and two broad Hubbard bands are seen in all curves. Upon hole doping, the lower Hubbard band shifts to cross the Fermi energy. Correspondingly, as shown in Fig. \ref{fig3}(b) for $p=0.05$, a narrow quasiparticle band emerges near $\omega=0$, which follows roughly the lower Hubbard band and moves from slightly below the Fermi energy around $\Gamma$ to above the Fermi energy around M. Clearly, this quasiparticle band is nothing but the hybridization band of the holons with the $\chi$ field via the spinons, and the dip can be viewed as an analog of the hybridization gap. Our results are similar to those from DMFT \cite{Pruschkea1996PhysicaB}, but provide a clearer physical picture.

\begin{figure}
	\includegraphics[width=9cm]{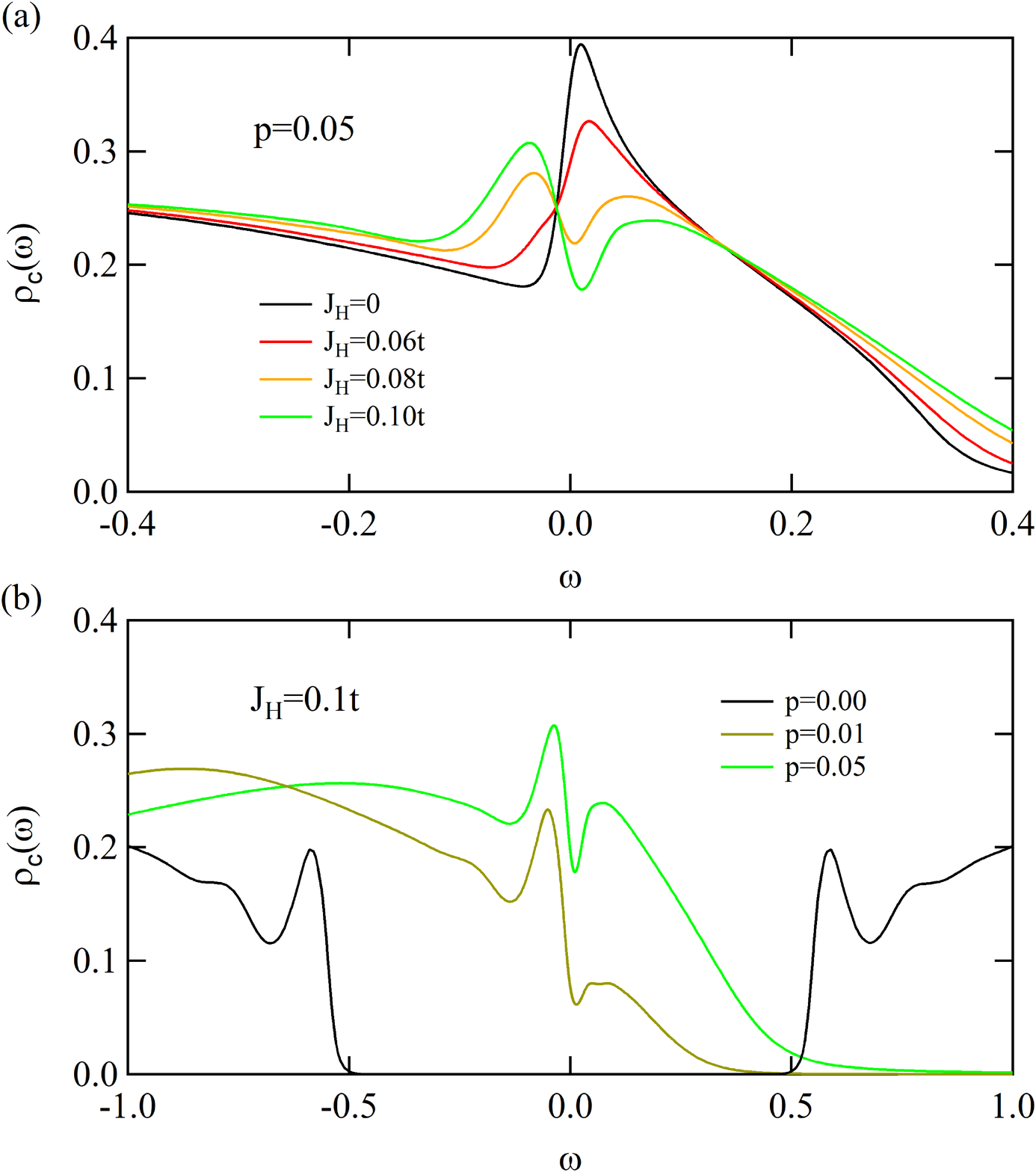}
	\caption{Comparison of the electron densities of states near $\omega=0$ for (a) different $J_H$ at fixed doping $p=0.05$ and (b) different doping at fixed $J_H=0.1t$ for $T=0.01$.}
	\label{fig4}
\end{figure}

For $J_H\neq 0$, four peaks appear in the spectra at half filling in Fig. \ref{fig3}(c). Upon hole doping, the lower Hubbard band again shifts across zero energy. From the shape of the spectra shown in Fig. \ref{fig3}(d), we may conclude that the left peak near $\omega=0$ is from the polaronic peak at the inner edge of the lower Hubbard band at half filling since it exhibits similar momentum dependence (inset) as in Fig. \ref{fig3}(c), while the right peak still follows the same behavior as for $J_H=0$, implying its hybridization origin. The two peaks are separated by an intermediate valley region, in correspondence with the pseudogap in the electron spectra for a finite $J_H$. It is important to compare our results with those of CDMFT \cite{Kyung2006CDMFTpseudogap}. At half filling, our observed four-peak structure in Fig. \ref{fig3}(c) is also captured by CDMFT in both paramagnetic and AFM solutions. On the other hand, some important differences may also be seen in Fig. \ref{fig3}(d). In the paramagnetic solution of CDMFT, the pseudogap occurs only near $(\pi,0)$ and there is a peak near $(\pi/2,\pi/2)$ at low energy. In our results, the pseudogap is seen at all momenta due to the local approximation of the self-energies. To reproduce the correct anisotropy of the pseudogap, one needs to go beyond the local approximation.

Figure \ref{fig4} summarizes how both peaks evolve with doping and $J_H$. In Fig. \ref{fig4}(a), the hybridization peak at $J_H=0$ is gradually suppressed and pushed to higher energy, while a new peak (the left peak) emerges and grows rapidly with increasing $J_H$, causing the pseudogap in between. 
Figure \ref{fig4}(b) shows the spectra with varying doping. Clearly, the left peak can be identified as the polaronic peak at $p=0$. The pseudogap therefore results from the interplay between the hybridization and polaronic mechanisms. From the spectra of $J_H=0$, it may also be viewed as the splitting of the quasiparticle peak due to polaron formation under AFM-correlated background. Hence, the spinons play two distinct roles: they act as an effective hybridization field between holons and the $\chi$-field to create electron quasiparticles, but in the meantime, they interact with holons to form polarons. The interplay of these two effects underlies the occurrence of the pseudogap phenomenon.

It may be helpful to comment briefly on the two approximations adopted in our calculations, namely the local approximation and the one-loop approximation. The local approximation ignores the momentum dependency of the self-energies in order to reduce the computation time. Our results show that it can already capture some major spectral features of the Mott state and the pseudogap by including nonlocal spin correlations through a Heisenberg mean-field term. However, the local approximation prevents us from producing the full momentum dependent features of the electron spectra such as the anisotropy of the pseudogap. It is therefore important to go beyond the local approximation for future investigations, possibly using the techniques employed previously for the Kondo lattice systems \cite{Nonlocal1,Nonlocal2,Nonlocal3,Nonlocal4}. The one-loop approximation ignores the effect of vertex corrections. A similar approximation has been used for the self-consistent Born approximation to study the spin-polaron problem in the $t$-$J$ model \cite{SpinPolaron1988,SpinPolaron1989,SpinPolaron1991,SpinPolaron1992}, where the vertex corrections were argued to be not crucial \cite{SpinPolaron1989,SpinPolaron1992}. It has also been applied to the Kondo problem and can yield important features including the Kondo hybridization \cite{HewsonBook}. In our case, the right peak of the pseudogap may be understood to arise from the Kondo effect of fermionic holons. Therefore, the two peaks associated with the pseudogap may survive beyond the one-loop approximation. Moreover, our results at $J_H=0$ agree well with those of DMFT, where all vertex corrections are included. At half filling, our method yields the correct Mott gap and the quantum Widom line over a wide $U$ range as in experiment or DMFT \cite{SlaveFermion2022HalfFilling}. In doped Mott insulators, our calculated resistivity at high temperatures and electron spectra also display similar behaviors as in DMFT. Both support the validity of our method beyond the one-loop approximation in this parameter region. On the other hand, for small $U$ or large doping  deep inside the metallic phase, we find some discrepancy with the Fermi liquid at low temperatures. This is not unexpected since the slave particle representation is not a suitable starting point for perturbative calculations to describe the Landau quasiparticles.

\section{Conclusion}
To summarize, we have generalized the recently developed slave fermion approach to study the doped Mott insulators in the one-band Hubbard and Hubbard-Heisenberg models away from half filling. In the absence of AFM correlations, we find a single sharp quasiparticle peak on the electron spectra due to the holon hybridization. Once AFM correlations are restored through a Heisenberg term, an additional polaronic peak emerges at slightly lower energy, causing the pseudogap in between around the zero energy. Our results are in good agreement with DMFT and CDMFT calculations. Our approach not only captures the key features of the electron spectra in doped Mott insulators but also gives a clearer picture of their origin. Our work may serve as a starting point for further development beyond the self-consistent one-loop and local approximations to achieve a better understanding of the doped Mott insulators.

\acknowledgements

This work was supported by the National Natural Science Foundation of China (Grants No. 11974397 and No. 12174429), the National Key R\&D Program of China (Grant No. 2022YFA1402203), and the Strategic Priority Research Program of the Chinese Academy of Sciences (Grant No. XDB33010100).

\appendix

\section{Self-consistent equations and numerical details}

We have used the following Dyson's equations with local self-energies:
\begin{eqnarray}
&&G_\chi(\bm k,\text{i}\omega_n)=\dfrac{\epsilon_{\bm k}-\mu_1}{1-(\epsilon_{\bm k}-\mu_1)\Sigma_\chi(\text{i}\omega_n)},\notag\\
&&G_s(\bm q,\text{i}\nu_m)=\dfrac{\gamma_s(-\text{i}\nu_m)}{\gamma_s(\text{i}\nu_m)\gamma_s(-\text{i}\nu_m)-|\Delta|^2\eta^2_{\bm  q}},\notag\\
&&G_{h}(\bm k,\text{i}\omega_n)=\dfrac{1}{\text{i}\omega_n-\lambda-\mu_2-U/2-\Sigma_h(\text{i}\omega_n)},\notag\\
&&G_{d}(\bm k,\text{i}\omega_n)=\dfrac{1}{\text{i}\omega_n-\lambda+\mu_2-U/2-\Sigma_d(\text{i}\omega_n)},\notag\\
&&G_c(\bm k,\text{i}\omega_n)=\dfrac{\Sigma_\chi(\text{i}\omega_n)}{1-(\epsilon_{\bm k}-\mu_1)\Sigma_\chi(\text{i}\omega_n)},
\end{eqnarray}
where $\gamma_s(\text{i}\nu_m)\equiv \text{i}\nu_m-\lambda-\Sigma_s(\text{i}\nu_m)$, $\epsilon_{\bm k}=-2t[\cos(k_x)+\cos(k_y)]$, $\eta_{\bm q}=2t[\sin(q_x)+\sin(q_y)]$. The corresponding local Green's functions are calculated using $G(\text{i}\omega_n)=N^{-1}\sum_{\bm k}G(\bm k,\text{i}\omega_n)$, where $N$ is the number of lattice sites. The sum over momentum can be done analytically, giving
\begin{widetext}
\begin{eqnarray}
&&G_\chi(\text{i}\omega_n)=\dfrac{1}{\Sigma_\chi(\text{i}\omega_n)}\left\lbrace\dfrac{2}{\pi [1+\mu_1\Sigma_\chi(\text{i}\omega_n)]} \cdot K\left(\dfrac{4t\Sigma_\chi(\text{i}\omega_n)}{1+\mu_1\Sigma_\chi(\text{i}\omega_n)}\right)-1 \right\rbrace,\notag\\
&&G_s(\text{i}\nu_m)=\dfrac{2}{\pi\gamma_s(\text{i}\nu_m)}\cdot K\left[\left(\dfrac{16t^2|\Delta|^2}{\gamma_s(\text{i}\nu_m)\gamma_s(-\text{i}\nu_m)}\right)^{1/2}\right],\notag\\
&&G_{h}(\text{i}\omega_n)=\dfrac{1}{\text{i}\omega_n-\lambda-\mu_2-U/2-\Sigma_h(\text{i}\omega_n)},\notag\\
&&G_{d}(\text{i}\omega_n)=\dfrac{1}{\text{i}\omega_n-\lambda+\mu_2-U/2-\Sigma_d(\text{i}\omega_n)},\notag\\
&&G_c(\text{i}\omega_n)=\dfrac{2\Sigma_\chi(\text{i}\omega_n)}{\pi[1+\mu_1\Sigma_\chi(\text{i}\omega_n)]}\cdot K\left(\dfrac{4t\Sigma_\chi(\text{i}\omega_n)}{1+\mu_1\Sigma_\chi(\text{i}\omega_n)}\right),
\end{eqnarray}
\end{widetext}
where $K(z)$ is the complete elliptic integral of the first kind:
\begin{eqnarray}
K(z)\equiv \int_0^{\pi/2}\dfrac{\text{d}\theta}{\sqrt{1-z^2\sin^2\theta}}.
\end{eqnarray}

In each step, we choose the gauge $A=-\text{i}|A|$ and determine the mean-field parameters $A\equiv\sum_\sigma\langle\sigma s_{j,\sigma}s_{i,-\sigma}\rangle$ and $\Delta\equiv |A|J_H/(2t)$ self-consistently by using
\begin{eqnarray}
A=\text{i}\dfrac{1}{2t}\dfrac{1}{N}\sum_{\bm q}\eta_{\bm q}\langle s_{-\bm q,\downarrow}s_{\bm q,\uparrow}\rangle,
\end{eqnarray}
in which
\begin{eqnarray}
\langle s_{-\bm q,\downarrow}s_{\bm q,\uparrow}\rangle&=&\dfrac{1}{\beta}\sum_m[-F_s(\bm q,\text{i}\nu_m)]\text{e}^{\text{i}\nu_m0^+},
\end{eqnarray}
where $F_s(\bm q,\text{i}\nu_m)\equiv -\langle s_{\bm q,\uparrow,m}s_{-\bm q,\downarrow,-m}\rangle$ is the anomalous Green's function of the spinon given by
\begin{eqnarray}
F_s(\bm q,\text{i}\nu_m)=\dfrac{[\text{i}AJ_H/(2t)]\eta_{\bm q}}{\gamma_s(\text{i}\nu_m)\gamma_s(-\text{i}\nu_m)-|\Delta|^2\eta^2_{\bm  q}}.
\end{eqnarray}
Here, the summation over Matsubara frequencies can be transformed into the integration over real frequency, and the sum over $\bm q$ can be done analytically. The final mean-field equation for $\Delta$ is
\begin{widetext}
\begin{align}
\dfrac{1}{J_H}=\dfrac{1}{4t^2|\Delta|^2}\int_{-\infty}^{\infty}\dfrac{\text{d}z}{\pi}n_B(z)\text{Im}\left\lbrace-1+\dfrac{2}{\pi}\cdot K\left[ \left(\dfrac{16t^2|\Delta|^2}{\gamma_s(z+\text{i}0^+)\gamma_s(-z-\text{i}0^+)}\right)^{1/2}\right]  \right\rbrace .
\end{align}
In real frequency, the self-energy equations are
\begin{eqnarray}
\Sigma''_\chi(\varepsilon)&=&\int\dfrac{\text{d}z}{-\pi}n_F(z+\varepsilon)G''_s(z)G''_d(z+\varepsilon)+\int\dfrac{\text{d}z}{-\pi}n_B(z)G''_s(z)G''_d(z+\varepsilon)\notag\\
&-&\int\dfrac{\text{d}z}{-\pi}n_F(z+\varepsilon)G''_s(-z)G''_h(-\varepsilon-z)-\int\dfrac{\text{d}z}{-\pi}n_B(z)G''_s(-z)G''_h(-\varepsilon-z),\\
\Sigma'_\chi(\varepsilon)&=&\int\dfrac{\text{d} z}{-\pi} n_B(z)G''_s(z)G'_d(z+\varepsilon)-\int\dfrac{\text{d} z}{-\pi} n_F(z)G''_d(z)G'_s(z-\varepsilon)\notag\\
&-&\int\dfrac{\text{d} z}{-\pi} n_B(z)G''_s(z)G'_h(z- \varepsilon)+\int\dfrac{\text{d} z}{-\pi} n_F(z)G''_h(z)G'_s(z+\varepsilon),\\
\Sigma''_s(\varepsilon)&=&-\int\dfrac{\text{d} z}{-\pi}n_F(z+\varepsilon)G''_\chi(z)G''_d(z+\varepsilon)+\int\dfrac{\text{d} z}{-\pi}n_F(z)G''_\chi(z)G''_d(z+\varepsilon)\notag\\
&-&\int\dfrac{\text{d} z}{-\pi}n_F(z+\varepsilon)G''_\chi(-z)G''_h(z+\varepsilon)+\int\dfrac{\text{d} z}{-\pi}n_F(z)G''_\chi(-z)G''_h(z+\varepsilon),\\
\Sigma'_s(\varepsilon)&=&\int\dfrac{\text{d}z}{-\pi} n_F(z)G''_\chi(z)G'_d(z+\varepsilon)+\int\dfrac{\text{d}z}{-\pi} n_F(-z)G''_d(-z)G'_\chi(-z-\varepsilon)\notag\\
&-&\int\dfrac{\text{d}z}{-\pi} n_F(-z)G''_\chi(-z)G'_h(z+\varepsilon)+\int\dfrac{\text{d}z}{-\pi} n_F(z)G''_h(-z)G'_\chi(z+\varepsilon),\\
\Sigma''_d(\varepsilon)&=&2\int\dfrac{\text{d} z}{-\pi}n_F(z-\varepsilon)G''_s(z)G''_\chi(\varepsilon-z)+2\int\dfrac{\text{d} z}{-\pi}n_B(z)G''_s(z)G''_\chi(\varepsilon-z),\\
\Sigma'_d(\varepsilon)&=&2\int\dfrac{\text{d} z}{-\pi} n_B(z)G''_s(z)G'_\chi (-z+\varepsilon)+2\int\dfrac{\text{d} z}{-\pi} n_F(z)G''_\chi(-z)G'_s(z+\varepsilon),\\
\Sigma''_h(\varepsilon)&=&2\int\dfrac{\text{d} z}{-\pi}n_F(z-\varepsilon)G''_s(z)G''_\chi(z-\varepsilon)+2\int\dfrac{\text{d} z}{-\pi}n_B(z)G''_s(z)G''_\chi(z-\varepsilon),\\
\Sigma'_h(\varepsilon)&=&-2\int\dfrac{\text{d} z}{-\pi} n_B(z)G''_s(z)G'_\chi(z-\varepsilon)+2\int\dfrac{\text{d} z}{-\pi} n_F(z)G''_\chi(z)G'_s(z+\varepsilon),
\end{eqnarray}
\end{widetext}
where $G'$ and $G''$ ($\Sigma'$ and $\Sigma''$) are the real and imaginary parts of the Green's functions $G$ (self-energies $\Sigma$), respectively.

All our calculations are carried out in real frequency using several different frequency grids simultaneously. The first grid contains points like $\{\pm10^{x}\}$ for $x$ from $1.5$ to $-3$ with step $0.01$, which covers the region $|\omega|/D>10^{-3}$. The second grid covers $(-10^{-3},10^{-3})$ uniformly with the step $2\times10^{-5}$. We also use some other grids to deal with the peak or edge features of the spectra. The Lorentzian broadening factor is set to $10^{-5}$, because the position of the spinon peak at $J_H=0$ and the spinon gap for $J_H>0$ are all very small at low temperature.

\end{document}